\documentclass{PoS}

\title{Do we know eventually $\bf p \, (e)$?}

\ShortTitle{Do we know eventually $p \, (e)$?}

\author{\speaker{B. K\"ampfer}%
         \thanks{also at TU Dresden, 01062 Dresden, Germany}\\
        Forschungszentrum Dresden-Rossendorf, PF 510119, 01314 Dresden, Germany\\
        E-mail: \email{b.kaempfer@fzd.de}}

\author{M. Bluhm, H. Schade, R. Schulze, D. Seipt\\
Forschungszentrum Dresden-Rossendorf, PF 510119, 01314 Dresden, Germany\\
}

\abstract{
A quasi-particle model is employed to derive from available lattice
QCD calculations an equation of state useable in hydrodynamical simulations
of the expansion stage of strongly interacting matter created in
ultra-relativistic heavy-ion collisions.
Various lattice results give an astonishing agreement of the pressure as a function
of energy density at large energy densities supposed the pseudo-critical
temperature is in the range $170 \pm 15$ MeV, while in the transition
region the equation of state is not yet well constrained. Therefore, one can construct
a family of equations of state by bridging the uncertain region from the uniquely
given high-energy density region part to a hadronic equation of state by suitable 
interpolation together with the extrapolation to non-zero baryon density
by means of the quasi-particle model. We present a series of tests of the model, 
discuss the chiral extrapolation and the role of Landau damping.
We also briefly sketch the path of cosmic matter in the early universe
in the phase diagram.}

\FullConference{Critical Point and Onset of Deconfinement - 4th International Workshop\\
July 9 - 13, 2007\\
Darmstadt, Germany}

\begin{document}

\section{Introduction}

The equation of state of strongly interacting matter is a central issue
for understanding and modelling the adiabatic path of cosmic matter of
the expanding universe (hot QCD), 
the expansion of matter created in ultra-relativistic heavy-ion
collisions (hot and medium-dense QCD), and compact stars (dense QCD).
Intimately related is the phase diagram of strongly interacting matter
with challenges like the nature and localization of the deconfinement
transition and the occurrence of a critical point on the phase border curve.
Ab initio calculations evaluating observables which quantify these notions
are still in progress. It is, therefore, opportune to employ at this stage
of insight in QCD appropriate models to interpolate and extrapolate
the various pieces of knowledge aiming at delivering, e.g., a useable equation
of state.

The tool employed here is the quasi-particle model developed in
\cite{QPM_Peshier}. We are going to address the question whether a unique
equation of state, applicable for ultra-relativistic heavy-ion collisions,
is at our disposal. Furthermore, we report on a few recent improvements
of the model.

Our paper is organized as follows.
In section 2 we review the quasi-particle model, which is exploited
in section 3 to derive an equation of state suitable for predictions
for heavy-ion collisions at RHIC and LHC energies. The possibility 
to extend the model by explicitly accounting for the critical point 
is addressed in section 4. Partial improvements 
(such as an attempt of chiral extrapolation, and inclusion of Landau
damping and other plasma excitations, and imaginary chemical potential)
are briefly reported in section 5. Finally, we
comment on the adiabatic path of strongly interacting matter in the
expanding universe (section 6).

\section{Quasi-particle model \label{QPM}}

The present basic version of the employed quasi-particle model is related
to QCD as follows \cite{MB_EPJC}:
(i) two-loop $\Phi$ functional which results in one-loop self-energies,
(ii) neglect of imaginary parts of self-energies (and, via Dyson's relation,
also in propagators) as well as neglect of (anti)plasmino and 
longitudinal gluon excitations, 
(iii) use of approximate energy ($\omega$) - momentum ($k$)
relations of quasi-particle excitations 
\begin{eqnarray} 
\label{gluon_dispersion}
\omega_T^2 &=& k^2 + m_\infty^2 , \quad
m_\infty^2 = \frac{1}{12}
\left( [2N_c + N_f] T^2 + \frac{N_c}{\pi^2} N_f\mu_q^2 \right) G^2(T,\mu_q),\\
\label{quark_dispersion}
\omega_q^2 &=& k^2 + m_q^2 + 2m_qM_+ + 2 M_+^2 , \quad
M_+^2 = \frac{N_c^2 -1}{16 N_c} \left(T^2 + \frac{\mu_q^2}{\pi^2}\right) G^2 (T, \mu_q)%
\end{eqnarray} 
for transverse gluons ($T$) and quarks ($q$), where
(iv) the effective coupling $G^2$ obeys Peshier's equation 
$a_T \frac{\partial G^2}{\partial T} +
a_{\mu_q} \frac{\partial G^2}{\partial \mu_q} + a_G G^2 =0$
with coefficients $a_{\mu_q, T, G}$ given in \cite{MB_EPJC,Bluhm_Diploma},
once $G^2(T, \mu_q =0)$ in the parametrization
\begin{eqnarray}
\label{eff_G}
G^2(T) = \left\{
\begin{array}{ll}
G^2_{\rm 2 loop} (\xi),\hspace*{4mm} \xi=\lambda \frac{(T-T_s)}{T_c}, &T \ge T_c
\\[3mm]
G^2_{\rm 2 loop}(T_c) + b (1 - \frac{T}{T_c}),&T < T_c
\end{array}
\right.
\label{effective_copuling}
\end{eqnarray}
is adjusted to lattice data. 
($G^2_{\rm 2 loop}$ is the two-loop running QCD coupling, however, with
argument $\xi$.)
The expressions for baryon and entropy density look
like the standard statistical integrals, however, with state-dependent
mass gaps $m_\infty^2$ and $M^2=m_q^2+2m_qM_++2M_+^2$ for gluons and 
quarks. For quarks, the existence of a mass gap 
related to $M_+$ emerges now from lattice simulations in the quenched 
approximation~\cite{Karsch_Kitazawa}. The dependence of the latter ones on 
the quark chemical potential $\mu_q$, beyond the dependence on the temperature $T$, 
is called the BKS effect in \cite{Shuryak}.

\begin{figure}[h]
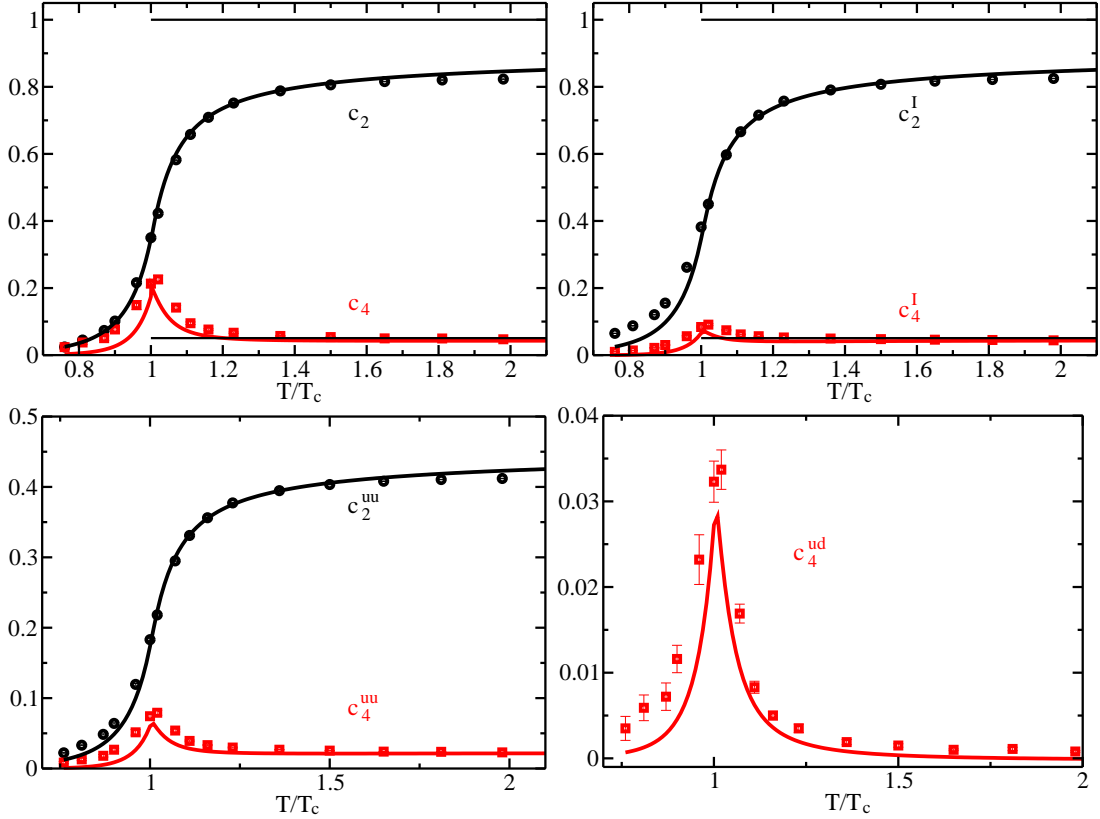
  %  
\includegraphics[scale=0.31,angle=0.,clip=]{c2c4_1.eps} %
\includegraphics[scale=0.31,angle=0.,clip=]{c2Ic4I.eps} \\ \hskip 1cm%
\includegraphics[scale=0.31,angle=0.,clip=]{c2uuc4uu.eps} %
\includegraphics[scale=0.31,angle=0.,clip=]{c2udc4ud_1.eps} %
\vskip -3mm
\caption{\label{susceptibilities} 
Taylor expansion coefficients for susceptibilities. 
Lattice QCD data from \cite{Allton}. Straight lines for $T > T_c$ depict
here and in the following the perturbative results.}
\end{figure}

The model has been tested against various sets of lattice QCD results
for $N_f$ flavors and $N_c = 3$ colors, see
\cite{QPM_Peshier,QPM_Bluhm}. A recent extension of the model towards
two independent chemical potentials $\mu_{u,d}$ allows the calculation of 
quark number and isovector susceptibilities as well as diagonal and
off-diagonal susceptibilities and their respective Taylor expansion coefficients
\begin{eqnarray}
\frac{\chi_q}{T^2} = 2 c_2 + 12 c_4 \left( \frac{\mu_q}{T} \right)^2 + \cdots,\\
\frac{\chi_I}{T^2} = 2 c_2^I + 12 c_4^I \left( \frac{\mu_I}{T} \right)^2 + \cdots,\\
\frac{\chi_{uu}}{T^2} = 2 c_2^{uu} + 12 c_4^{uu} \left( \frac{\mu_q}{T} \right)^2 + \cdots,\\
\frac{\chi_{ud}}{T^2} = 2 c_2^{ud} + 12 c_4^{ud} \left( \frac{\mu_q}{T} \right)^2 + \cdots,
\end{eqnarray}
being second-order derivatives of the grand thermodynamical potential. 
The light-quark chemical potentials are decomposed as
$\mu_q = \frac12 (\mu_u + \mu_d)$ and $\mu_I = \frac12 (\mu_u - \mu_d)$, where 
$\mu_I$ denotes the isospin chemical potential. 
A comparison with lattice QCD results is displayed in Fig.~\ref{susceptibilities}.
This additional test provides further confidence in the model.

\section{Hydrodynamics for RHIC and LHC}

\begin{figure}[h]
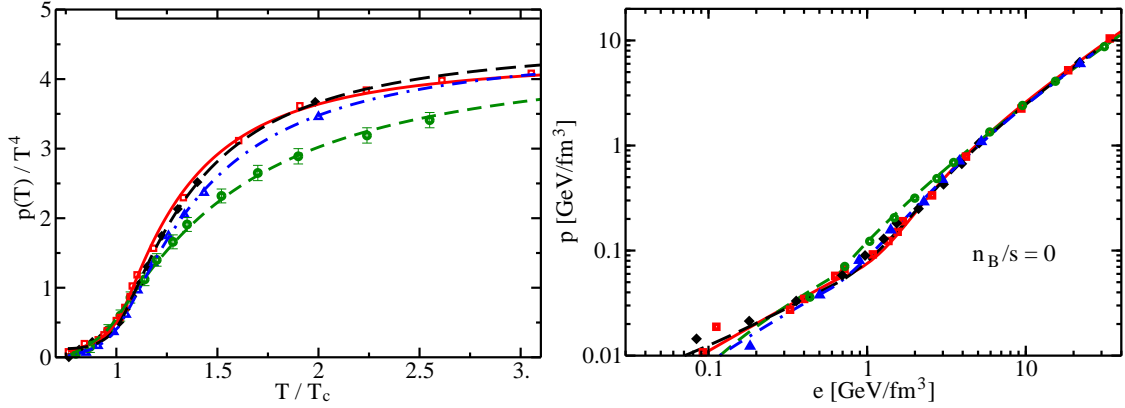
  %  
\includegraphics[scale=0.297,angle=0]{pres6_1.eps} %
\includegraphics[scale=0.31,angle=0]{EoS9_1.eps} %
\vskip -3mm
\caption{\label{EoS} 
Left panel: Scaled pressure as a function of scaled temperature
(quasi-particle model [curves] adjusted to various selected
lattice QCD data [symbols] from \cite{Kar2,Bernard,Fodorlat3}).
Right panel: Resulting equation of state in the form of pressure $p$ vs.\ energy density
$e$ when using the scale $T_c = 170$ MeV. The curves are robust against 
variations in $T_c$ for small and large $e$. Only in the transition region 
$1$ GeV/fm$^3 \le$ e $\le 5$ GeV/fm$^3$ noticeable differences of at most 
$20 \%$ arise when modifying $T_c$ by $\pm 10$ MeV.}
\end{figure}

Due to various differences of the implementation of QCD on a space-time
grid to evaluate the equation of state for $N_f = 2 + 1$ flavors, 
some differences are obvious,
as demonstrated in the left panel of Fig.~\ref{EoS}.
The surprise however is that a translation of these differing lattice results for
the scaled pressure $p$ as a function of the scaled temperature into the
relation of pressure as a function of energy density $e$ by means of the 
above quasi-particle model provides a unique
equation of state at large energy densities. In this respect one may 
arrive at the conclusion that the equation of state $p(e)$ in the form pressure
vs.\ energy density is eventually at our disposal for large values of $e$. 
As the calculations exhibited
in the left panel of Fig.~\ref{EoS} are for $\mu_q = 0$ we use again our
quasi-particle model to supplement the baryon density dependence.
The successful tests of the baryon density dependence of two-flavour QCD in
\cite{QPM_Bluhm} and the in the previous section provide confidence 
in the reliable extrapolation to $\mu_q > 0$. 

\vskip 6mm

The transition region, corresponding to temperatures of the oder of the
pseudo-critical temperature, $T \sim T_c$,
however, seems not yet to be settled. We construct, therefore, a family
of equations of state by interpolating from the unique high-density QCD part to a
low-density hadron resonance part. The resulting equation of state is
used for the hydrodynamical calculation of transverse momentum ($p_\perp$)
spectra and azimuthal asymmetry $v_2$ 
of various hadron species for RHIC and future LHC energies.
Keeping initial and freeze-out conditions fixed, the spectra and $v_2(p_\perp)$
show some dependence on the shape of $p(e)$ in the region around $T_c$.
This may enable further constraints on the equation of state by comparison
with experimental data from RHIC, a programme already pursued by \cite{Pasi}.
For results and details the interested reader is referred to \cite{UH,MB_CERN}.    

\section{Including the critical point}

The change of the slope of the employed effective coupling (\ref{effective_copuling}) 
at $T_c$ is primarily responsible for the pronounced structures observed in 
$c_4$, $c_4^I$, $c_4^{uu}$ and $c_4^{ud}$ in Fig.~\ref{susceptibilities}. 
While the rise for $T\ge T_c$ close to $T_c$ is dictated by the perturbative behavior 
of $G^2$, the decrease for $T < T_c$ is due to the conversion into the linear 
temperature dependence. Such a change in the curvature behavior 
was also found by solving Dyson-Schwinger equations in Coulomb gauge 
\cite{Reinhardt}. This behavior may be interpreted
as an indication of some criticality in agreement with lattice QCD results, 
nonetheless, one can further extend the model by including explicitly a singular part
which belongs to the universality class of the three-dimensional Ising model.
In such a way the conjectured QCD critical point can be modelled without
destroying the agreement with the available lattice QCD results \cite{Allton}.
Such a phenomenological procedure, first proposed in \cite{Yuki},
is described in \cite{CP_POS_QM05}. It offers the opportunity to study various
observables within a hydrodynamical framework for matter states in the vicinity
of the critical point, similar to first investigations along this line
in \cite{Frankfurt}. 

\section{Recent developments}

The severe approximations described in beginning of section \ref{QPM} are matter of ongoing
investigations. Here we describe some of such studies. 

\subsection{Chiral extrapolation}

\begin{figure}[h]%
\center
\includegraphics[scale=0.28,angle=0]{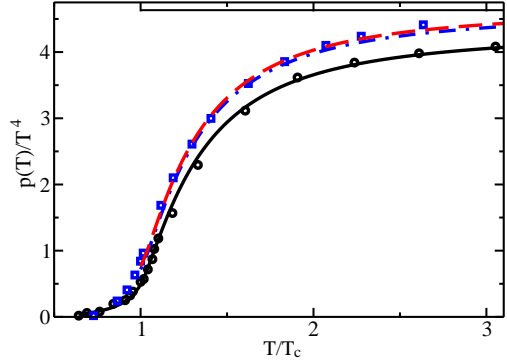} %
\vskip -3mm
\caption{\label{chiral_extrapolation} 
Scaled pressure as a function of scaled temperature. 
Symbols depict lattice QCD data for $2+1$ flavors from \cite{Kar2}
(full circles, $m_{u,d} = 0.4 T$, $m_s = T$) and 
\cite{Schmidt_2007} (squares, $m_\pi = 220$ MeV). 
The black solid curve employs quark masses as on the lattice,
the red dashed curve is for the chiral limit, while the blue dash-dotted curve
uses $m_{u,d} = 0.024 T$ MeV and $m_s = 0.24 T$ MeV.}
\end{figure}

Lattice performances require still fairly large quark masses.
Given the form of the approximated quark dispersion relation 
(\ref{quark_dispersion}) with lattice mass parameter $m_q$ 
one can attempt to perform a chiral extrapolation by putting $m_q \to 0$. 
The result of such a procedure is exhibited in Fig.~\ref{chiral_extrapolation}.
The semi-quantitative agreement with new lattice QCD results with
smaller quark masses supports the idea that an implicit dependence of
$G^2$ on quark masses is weak.

A calculation of the quasi-particle dispersion relations based on 
one-loop self-energies with finite quark masses in Feynman gauge
\cite{D_Seipt} renders the above simplified expressions 
(\ref{gluon_dispersion}) and (\ref{quark_dispersion}) at $\mu_q=0$ into
\begin{eqnarray}
\label{gluon_disp}
\omega_T^2 &=& k^2 + m_\infty^2 , \quad 
m_\infty^2 = \frac 16 G^2 T^2 \left(N_c  + \frac12 \sum_q {\cal I} (\frac{m_q}{T}) \right),\\
\label{quark_disp}
\omega_q^2 &=& k^2 + m_q^2 + 2 M_+^2 , \quad 
M_+^2 = \frac16 G^2 T^2 \left( \frac23 + \frac13 {\cal I} (\frac{m_q}{T} ) \right)
\end{eqnarray}
with the asymptotic representation for small values of $m_q/T$ 
\begin{eqnarray}
{\cal I} (x) &=& 1 + a_2 x^2 + a_L x^2 \log x^2 + a_4 x^4 + \cdots,\\
a_2 &=& - \frac{3}{\pi^2} ( \log \pi + \frac12 - \gamma_E),\\
a_L &=& \frac{3}{2 \pi^2},\\
a_4 &=& - \frac{21}{16 \pi^4} \xi (3).
\end{eqnarray}
The asymptotic masses $m_\infty^2$ and $M_\infty^2=m_q^2+2M_+^2$ are found 
to be gauge invariant quantities \cite{D_Seipt}. 
(\ref{gluon_disp}) and (\ref{quark_disp}) as well as 
improved dispersion relations offer the possibility of sound 
chiral extrapolations. \vskip 7mm

\subsection{Landau damping}

While the quasi-particle model as outlined in section \ref{QPM}
describes fairly well the Taylor expansion coefficients of \cite{Allton},
thus establishing the correct $\mu_q$ dependence for small $\mu_q$,
it exhibits an ostensible ambiguity at larger $\mu_q$: By solving 
Peshier's equation to determine $G^2(T,\mu_q)$, the 
characteristic curves emerging from the vicinity of $T_c$ cross each other
at larger $\mu_q$. As indicated in \cite{Romatschke}, 
including the imaginary parts of HTL/HDL self-energies cures this
unpleasant feature. A detailed study \cite{R_Schulze} shows that
the negative longitudinal gluon and (anti)plasmino contributions to 
the entropy density allow for a better adjustment of the model 
to lattice QCD data \cite{Schmidt_2007} leading to less crossings. However, 
the complete cure of crossings is due to Landau damping terms 
within Peshier's equation. 

\subsection{Imaginary chemical potential}

\begin{figure}[h]
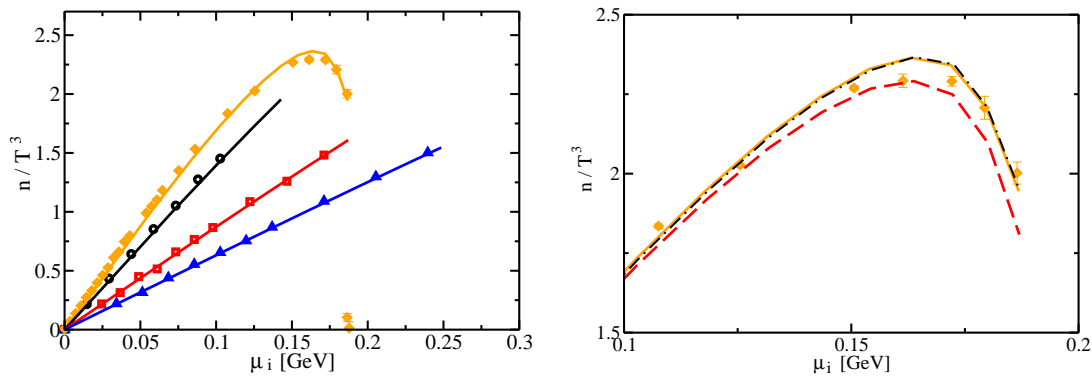
% 
\includegraphics[scale=0.28,angle=0.]{dens7_4.eps}%
\hskip 5mm
\includegraphics[scale=0.28,angle=0]{dens7_6_1.eps}%
\caption{\label{MPL_density} 
Left: scaled quark number density as a function of imaginary chemical potential 
for various temperatures. Symbols denote lattice QCD data from \cite{LPM_imag_mu} 
for $T=1.1,1.5,2.5,3.5\, T_c$ (diamonds, circles, squares and triangles, 
respectively). Right: test of the BKS effect for $T=1.1\, T_c$ as explained in 
the text.}
\end{figure}

The notorious sign problem of the fermionic determinant is
avoided for a purely imaginary chemical potential 
(see \cite{imaginary_mu} for recent reviews), 
where QCD recovers the center symmetry giving rise to the Roberge-Weiss
periodicity \cite{Roberge-Weiss}. Our quasi-particle model, as described
in section \ref{QPM} can be applied accordingly by the replacement of 
baryo-chemical potential $\mu_q \to i \mu_i \equiv \mu_B/3$. 
This replacement flips signs at a few important points
in the equations for the thermodynamic quantities, the self-energies and
the Peshier equation. For instance, $m_\infty^2$ and $M_+^2$ in 
(\ref{gluon_dispersion}) and (\ref{quark_dispersion}) render to 
\begin{eqnarray} 
\label{dispglue}
m_\infty^2 & = & \frac{1}{12}
\left( [2N_c + N_f] T^2 - \frac{N_c}{\pi^2} N_f\mu_i^2 \right) G^2(T, i\mu_i),\\
\label{dispquark}
M_+^2 & = & \frac{N_c^2 -1}{16 N_c} \left(T^2 - \frac{\mu_i^2}{\pi^2}\right) G^2 (T, i\mu_i)\,.%
\end{eqnarray} 
This comprises an additional sensible test,
even if the truncation of QCD by the chosen $\Phi$ functional, the
approximated HTL/HDL self-energies and the neglect of imaginary parts 
in the self-energies discards the center symmetry of full QCD
with imaginary chamical potential.
A comparison with lattice QCD results is displayed in Fig.~\ref{MPL_density}. 
In particular, the importance of the BKS-effect \cite{Shuryak,LPM_imag_mu} 
for the successful description of the observed pattern in the quark number 
density can be studied as depicted in the rigth panel of 
Fig.~\ref{MPL_density}. 
For example, one could neglect the terms explicitly depending on $\mu_i$ in the 
quasi-particle dispersion relations with (\ref{dispglue}) and (\ref{dispquark}) 
which enter the thermodynamic expression of the quark number density 
but leaving Peshier's equation unchanged. This renders the found results only 
at larger values of $\mu_i$ (dashed line). Thermodynamic consistency, nonetheless, 
requires in addition corresponding changes in Peshier's equation resulting 
in negligible deviations (dash-dotted line) from the results shown in the 
left panel of Fig.~\ref{MPL_density}. Thus, one may conclude that an explicit 
$\mu_i$-dependence of the quasi-particle's effective masses is not 
necessarily required for describing these lattice QCD results 
\cite{LPM_imag_mu}; instead the $\mu_i$-dependence of $G^2$ matters. 

\section{Strongly interacting matter in the expanding universe}

\begin{figure}[h]  %
%\vskip -15cm
~\centering
\includegraphics[width=0.43\textwidth]{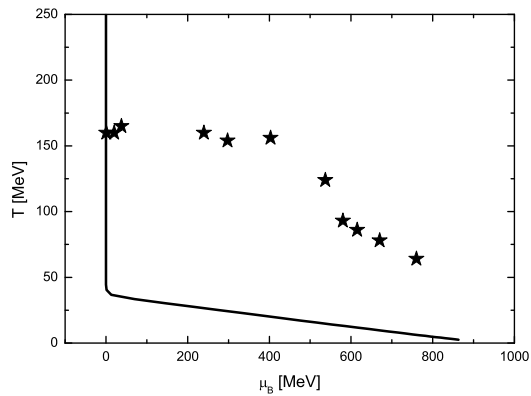} %
%\vskip -1mm
\caption{\label{universe} 
Adiabatic path of cosmic matter for inverse specific entropy of 
$2.3 \times 10^{-11}$, assuming about
10 degrees of freedom (photons, standard model neutrinos, electrons)
for the entropy density.  
The asterisks depict chemical freeze-out points from \cite{hadro_chemistry}
(table 2--upper part, and a LHC estimate mentioned in the text there).}
\end{figure}

The WMAP data on the fluctuations of the cosmic microwave background
deliver a value of $6.1 \times 10^{-10}$ ($\Lambda$CDM model
and 3-year WMAP-data-only \cite{WMAP}) for the ratio of
baryons to photons. Assuming adiabaticity and baryon conservation
this translates into $2.3 \times 10^{-11}$ for the inverse of the specific
entropy of baryons. The corresponding adiabatic path is displayed in the
temperature--baryo-chemical potential plane in Fig.~\ref{universe} 
(for plots in log scales cf.~\cite{SQM04,H_Schade}).
A similar path has been reported in \cite{PBM_universe}, where also
charge neutrality and lepton conservation are implemented.
Intriguing is the sharp turn from a region of small baryo-chemical potential,
$\mu_B /T \ll 1$,
to large one, $\mu_B / T \gg 1$,
at temperature scale slightly below 50 MeV.
On the displayed path the time varies from 1~$\mu$sec to 0.1~sec, and
the universe is radiation dominated. 
At $T > T_c$ the strongly interacting (deconfined)
matter dominates over the electro-weak matter, while at $T < T_c$
the electro-weak matter dominates for a long time until recombination. 
''Dominating'' means here that the contribution to energy density, pressure
or entropy density exceeds the other contributions.

\vskip 4mm
  
\section{Summary}

In summary we survey the status of our quasi-particle model
and compare it with lattice QCD data. We argue that the equation
of state in the form pressure as a function of energy density, $p(e)$,
is eventually known at large energy densities and small baryo-chemical
potential. 

\subsection*{Acknowledgements}
The authors thank 
P. Braun-Munzinger, F. Karsch, E. Laermann, M.~P. Lombardo
for useful discussions.
The work is supported by BMBF 06DR136, GSI DRKAM, and EU I3HP.

\end{document}